%% file: CMSB_T1D_Behavior_Detection_2019.tex
\documentclass[runningheads]{llncs}
\usepackage{graphicx}
\usepackage{tabularx}
\usepackage{xcolor}
\usepackage{subfigure}
\usepackage{amssymb}
\usepackage{amsmath}

\newcommand{\startpara}[1]{{\vskip1pt\noindent{\bf #1}}}

\def\until{\mathbf{U}}
\newcommand{\sectref}[1]{Section~\ref{#1}}
\newcommand{\figref}[1]{Figure~\ref{#1}}
\newcommand{\tabref}[1]{Table~\ref{#1}}

\begin{document}

\title{A Logic-Based Learning Approach \\ to Explore Diabetes Patient Behaviors}
%
%
%
\author{Josephine Lamp \inst{1} \and
Simone Silvetti \inst{3} \and
Marc Breton \inst{2} \and
Laura Nenzi \inst{4} \and
Lu Feng \inst{1}}
\authorrunning{J. Lamp et al.}
%
\institute{Department of Computer Science, University of Virginia, Charlottesville, VA, USA \email{jl4rj@virginia.edu, lu.feng@virginia.edu}\and Center for Diabetes Technology, University of Virginia, Charlottesville, VA, USA \email{mb6nt@virginia.edu}\and
Esteco S.p.A., Trieste, Italy \email{simone.silvetti@gmail.com}\and
University of Trieste, Trieste, Italy \email{lnenzi@units.it}}

\maketitle              
\begin{abstract}
Type I Diabetes (T1D) is a chronic disease in which the body's ability to synthesize insulin is destroyed.
It can be difficult for patients to manage their T1D, as they must 
control a variety of behavioral factors that affect glycemic control outcomes. 
In this paper, we explore T1D patient behaviors using a Signal Temporal Logic (STL) based learning approach. 
STL formulas learned from real patient data characterize behavior patterns that may result in varying glycemic control.
Such logical characterizations can provide feedback to clinicians and their patients about behavioral changes that patients may implement to improve T1D control. 
We present both individual- and population-level behavior patterns learned from a clinical dataset of 21 T1D patients.



\end{abstract}
\input{introduction}
\input{preliminaries}
\input{methodology}
\input{results-individual}
\input{results-population}
\input{relatedwork}
\input{conclusion}

\section*{\centering Acknowledgements}
The authors would like to graciously thank the UVA Center for Diabetes Technology for providing the clinical datasets and Basak Ozaslan, Jack Corbett, Jonathan Hughes and Dr. Jos\'e Garc\'ia-Tirado for their clinical insights and valuable discussions. 
Research partially supported by the Austrian National Research Networks RiSE/ShiNE (S11405) and ADynNet (P28182) of the Austrian Science 
Fund (FWF).

\section*{\centering Appendix}\label{sec-appendix}

\begin{table}[htb]
\caption{Accuracy Rates for Repeated Rules}\label{table-mcr-temp-rules}
\centering
\begin{tabular}{|c|c|c|c|c|} \hline
\textbf{Patient} & \textbf{CGM \%} & \textbf{HR \%} & \textbf{Basal \%} & \textbf{Bolus \%}\\ \hline
1 & 88.61 & 51.25 & 86.94 & 90 \\ \hline
2 & 93.88 & 97.05 & 93.24 & 100\\ \hline
3 & 90.58 & 77.02 & 100 & 93.72 \\ \hline
4 & 96.10 & 56.25 & 98.48 & 98.48 \\ \hline
5 & 87.10 & 94.17 & 100 & 95.28 \\ \hline
6 & 88.19 & 51.25 & 100 & 97.92 \\ \hline
7 & 93.61 & 83.55 & 96.26 & 94.40 \\ \hline
8 & 87.08 & 78.89 & 99.72 & 100 \\ \hline
9 & 88.09 & 88.78 & 100 & 93.77 \\ \hline
10 & 95.45 & 97.47 & 100 & 95.36 \\ \hline
11 & 86.76 & 86.46 & 100 & 87.05 \\ \hline
12 & 93.75 & 55.17 & 100 & 95.63 \\ \hline
13 & 95.59 & 61.25 & 100 & 96.12 \\ \hline
14 & 94.40 & 79.33 & 96.38 & 100 \\ \hline
15 & 86.86 & 93.50 & 88.24 & 91.29 \\ \hline
16 & 89.38 & 75.47 & 90.90 & 89.13 \\ \hline
17 & 87.38 & 100 & 100 & 93.07 \\ \hline
18 & 89.54 & 71.09 & 90.66 & 89.65 \\ \hline
19 & 90.29 & 63.38 & 90.15 & 89.88 \\ \hline
20 & 89.54 & 62.43 & 91.11 & 89.99 \\ \hline
21 & 86.86 & 66.99 & 89.88 & 88.35 \\ \hline
\end{tabular}
\end{table}

%
%
\bibliographystyle{splncs04}
\bibliography{references}

\end{document}

%% file: introduction.tex
\section{Introduction}\label{sec-intro}
\begin{figure}[t]
\centering
\includegraphics[scale=0.28]{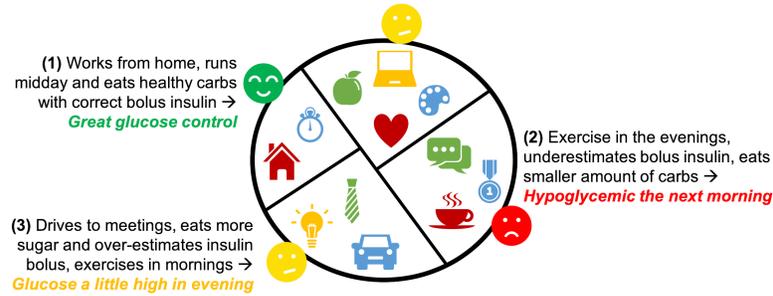}\label{fig-behavior-ex}
\caption{Hypothetical patient behaviors resulting in different glycemic control outcomes.} 
\vspace{-15pt}
\end{figure}

Type I Diabetes (T1D) is a chronic disease in which the body's ability to synthesize insulin is destroyed, as the patient's immune system attacks the insulin-producing cells of the pancreas~\cite{cdcT1d2018}. Insulin is an important hormone used by cells to absorb glucose for energy production. 
425 million people worldwide have Diabetes (Type I and Type II), including 1,106,500 children and adolescents living with T1D~\cite{idfatlas2017}. 
Intensive insulin therapy effectively reduces the risk of long-term complications of T1D (such as 
nerve or kidney damage) in which patients are required to inject or infuse insulin throughout the day to replace the normal pancreas function. Unfortunately, this means the burden of managing T1D 
falls to patients as they are 
required to manage a variety of behavioral factors (e.g., insulin injection, exercise, eating) that affect T1D.
Studies have found that such factors affect a patient's overall glycemic control: e.g., exercise may lower blood glucose values 
while 
carbs from meals increase blood glucose levels~\cite{american201913,Riddell2017}. 
\figref{fig-behavior-ex} shows a set of hypothetical patient behaviors that may result in varying glycemic control. For example, on days when 
a patient exercises in the evening and underestimates the insulin absorption amount, they may have poor glycemic control (hypoglycemia) the next morning. 
Characterization of these behaviors can be used by clinicians to counsel their patients on strategies to optimize glycemic control using predictive recommendations (e.g., if you exercise late at night, make sure you eat a snack before you go to bed to avoid morning hypoglycemia). However, it is challenging to accurately identify T1D patient behavior patterns due to inherently messy patient data and the individual variability of patient behavior and physiology. 

In this paper, we present a logic-based learning approach to address these challenges and explore T1D patient behaviors. Our approach 
takes advantage of the expressiveness and explainability of Signal Temporal Logic (STL)~\cite{maler2004monitoring} and 
uses STL learning~\cite{nenzi2018robust}
to learn a set of STL formulas that characterize both individual- and population-level T1D patient behaviors.  
We argue that STL is a suitable representation of patient behavior patterns, because it can capture the temporal relations of Diabetes patient actions and glycemic outcomes.
In addition, STL formulas are easily explainable to clinicians and patients. 
We apply our approach to learn STL formulas representing T1D patient behaviors from a clinical dataset including a variety of patient physiological and behavioral data,
such as Continuous Glucose Monitors (CGM) sensor readings, heart rate, step count and activity intensity recorded by Fitbit, insulin pump injection records, self-reported meals and blood glucose finger pricks (SMBG).  
We envision that the learned STL formulas can provide clinically-relevant insights for 
clinicians and patients to develop behavioral change strategies to improve glycemic control.

The rest of the paper is organized as follows:
\sectref{sec-prelim} introduces the background of STL and learning techniques. 
\sectref{sec-methods} describes our approach to learn STL formulas for characterizing T1D patient behaviors. 
\sectref{sec-res-individual} and \sectref{sec-res-pop} present our key findings about individual- and population-level patient behaviors, respectively. 
\sectref{sec-relatedwork} summarizes related work, and 
\sectref{sec-conclusion} draws conclusions and discusses future research directions.

%% file: preliminaries.tex
\section{Preliminaries}\label{sec-prelim}
In this section, we briefly introduce background on Signal Temporal Logic (STL) and STL learning techniques.
Formally, the syntax of an STL formula $\varphi$ is defined as follows:
$$\varphi ::= \mu \mid \neg \varphi \mid \varphi \land \varphi \mid \square_{(u,v)}\varphi \mid \Diamond_{(u,v)}\varphi \mid \varphi\space\until_{(u,v)}\space\varphi, $$
where $\mu$ is a signal predicate in the form of $g(\tau) > 0$ with a signal variable $\tau \in \mathcal{X}$ and function $g: \mathcal{X} \to \mathbb{R}$. 
The temporal operators $\square$, $\Diamond$, and $\until$ denote ``always'', ``eventually,'' and ``until'', respectively.
The bounded interval $(u,v)$ denotes the time interval of temporal operators and can be omitted if the interval is $[0,+\infty)$. 
For example, we can specify a diabetes management rule 
``continuous glucose monitoring signal should always be between 70 and 180"~\cite{young2018damon}
using a STL formula 
$\square(cgm \geq 70 \land cgm \leq 180)$.

\begin{figure}[t]
\centering
\subfigure[][]{%
\label{fig-beh-discriminate}%
\includegraphics[scale = 0.3]{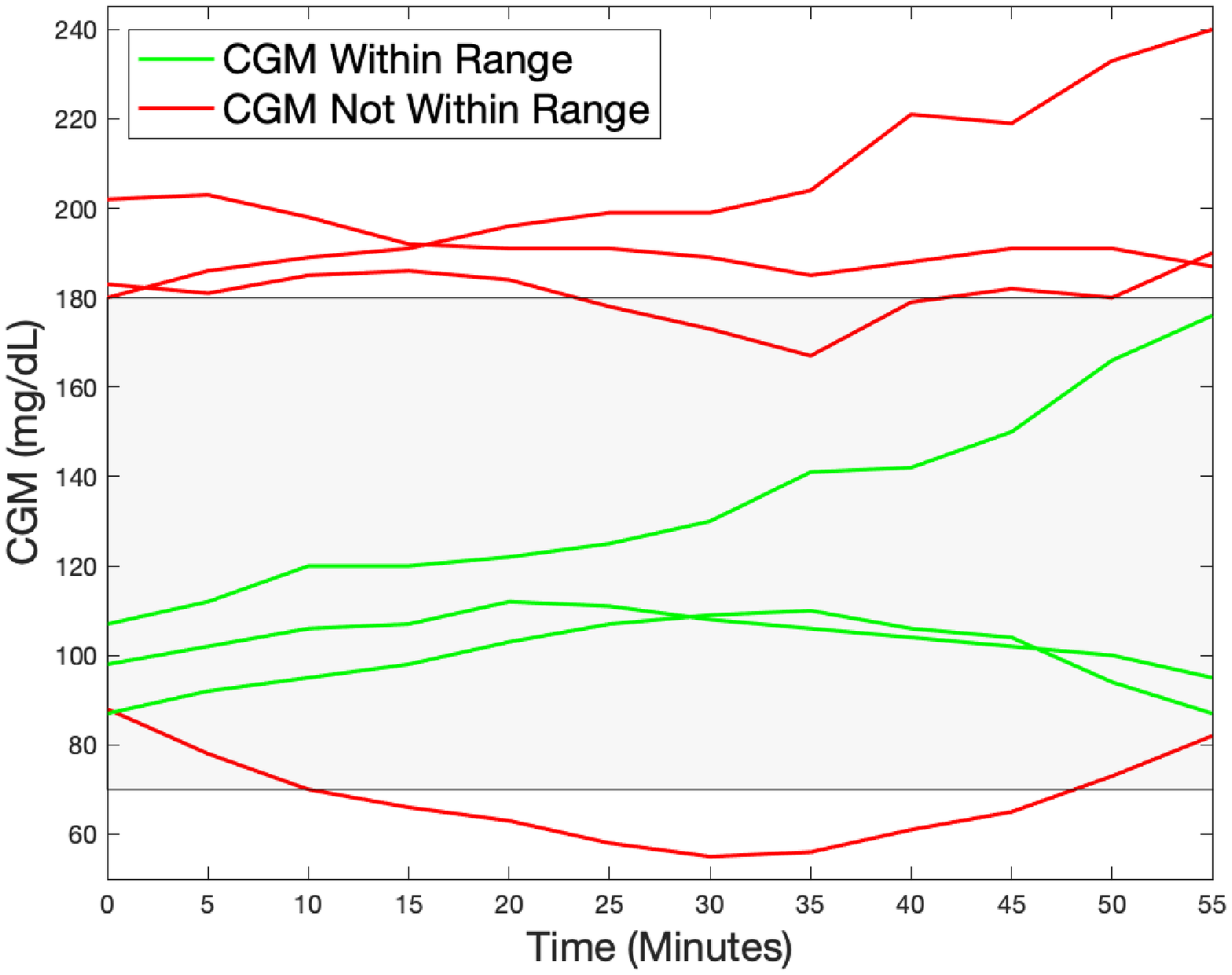}}%
\hspace{1pt}%
\subfigure[][]{%
\label{fig-pt-labels}%
\includegraphics[scale = 0.3]{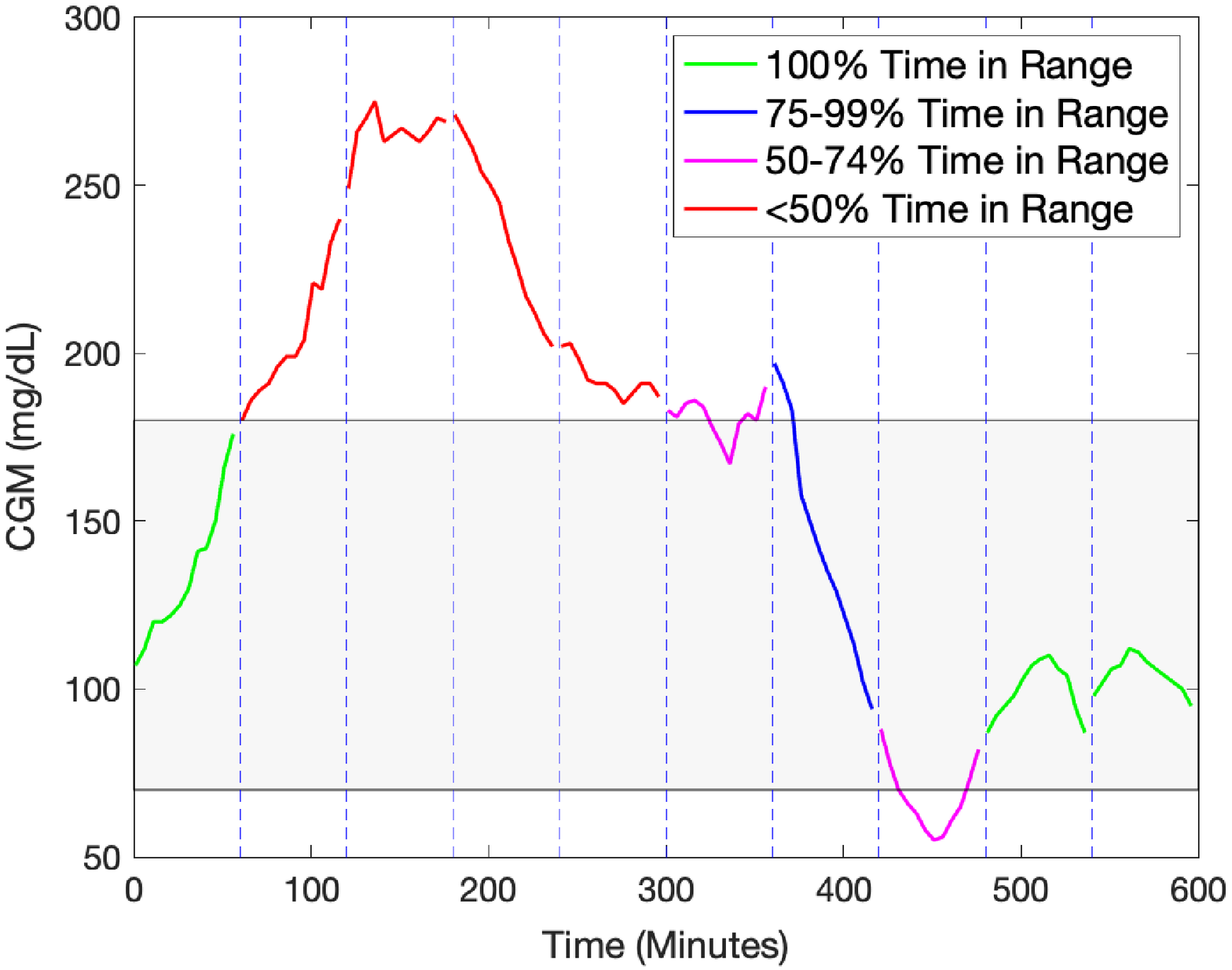}}%
\caption{\textbf{\subref{fig-beh-discriminate}} Example CGM trajectories that satisfy (green trajectories) or violate (red trajectories) the STL formula 
$\square(cgm \geq 70 \land cgm \leq 180)$.
\textbf{\subref{fig-pt-labels}} An example illustrating the labeling mechanism of patient data. The CGM trajectory is chopped into several one-hour chunks divided by the vertical dashed blue lines. Each chunk is assigned with one of the four labels based on the percentage of time that the CGM value is within the target grey region.}%
\vspace{-15pt}
\end{figure}

The satisfaction of a formula is verified over a signal trajectory. For example, the formula $\square(cgm \geq 70 \land cgm \leq 180)$ can be verified over the time series of CGM signals 
shown in \figref{fig-beh-discriminate}. 
STL considers two different semantics (Boolean and quantitative) to describe the satisfaction of a formula.
The Boolean semantics checks if a trajectory satisfies a STL formula.
For example, some CGM signals shown in \figref{fig-beh-discriminate} violate the STL formula $\square(cgm \geq 70 \land cgm \leq 180)$ because their CGM values go under 70 or above 180.
The quantitative semantics returns a real-valued \textit{robustness metric} that can be interpreted as a measure of the satisfaction~\cite{Deshmukh:2017}.  Signal trajectories exhibiting weakening robustness with respect to a given property can be said to be moving toward a state of violation.  
We refer to~\cite{BartocciBS14,BufoBSBLB14,Deshmukh:2017,Donze:2010:RST:1885174.1885183,FAINEKOS20094262} for a more detailed description of STL and its semantics. 

STL learning provides techniques to infer STL formulae and parameters from signal trajectories. STL learning goes beyond property specification and allows for the automated identification of interesting behaviors that may not initially be apparent to the human eye. 
Nenzi et. al.~\cite{nenzi2018robust} present a STL learning method that learns the best set of STL formulas to discriminate between
a two-label dataset of trajectories (e.g., regular and anomalous). 
This method uses a bi-level optimization process: it learns the STL formula structure using a discrete optimization of a genetic algorithm, and then synthesizes the parameters for the formulas using the Gaussian Process Upper Confidence Bound algorithm. We expand upon the STL learning tool developed in~\cite{nenzi2018robust} to learn STL formulas representing T1D patient behaviors.
However, since our clinical dataset has four labels (shown in \figref{fig-pt-labels}) rather than two labels, we need to adapt the tool for our problem. 
In addition, our goal is not to learn STL formulas that best discriminate between data with different labels. Instead, we are interested in learning STL formulas that can characterize patient behaviors that fall under the same label. 
We present our approach of learning STL formulas from T1D patient behaviors in the next section.

%% file: methodology.tex
\section{Methodology}\label{sec-methods}


We first describe the clinical dataset, then present our approach of learning individual- and population-level patient behaviors as illustrated in \figref{fig-methods}.

\subsection{Clinical Dataset Description}
Our dataset was collected during the observation period leading up to inpatient clinical trials in 2016-2017 at the Center for Diabetes Technology at the University of Virginia. The dataset contains 21 patients, ages ranging from 17 to 55, with an average age of 36 $\pm$ 10.4. Each patient has about 2 months of consecutive data. 
The data includes blood glucose readings from a Continuous Glucose Monitor (recorded in a variable named CGM), different types of insulin injections called boluses (total bolus, meal bolus, basal bolus, and correction bolus), meal carbs, patient-recorded blood glucose values from a finger prick (SMBG) and recordings of hypoglycemia (SMBG-Hypo). The data also contains exercise data recorded from a Fitbit including Heart Rate (HR), step count, calories, distance (in miles), and a Fitbit calculated activity level (in range of 1 to 4, with 1 being equivalent to little activity, and 4 being equivalent to intense activity.) 


\subsection{STL Learning for Individual Patient Behaviors}\label{sec-methods-individual}
\begin{figure}[t]
\centering
\includegraphics[scale=0.29]{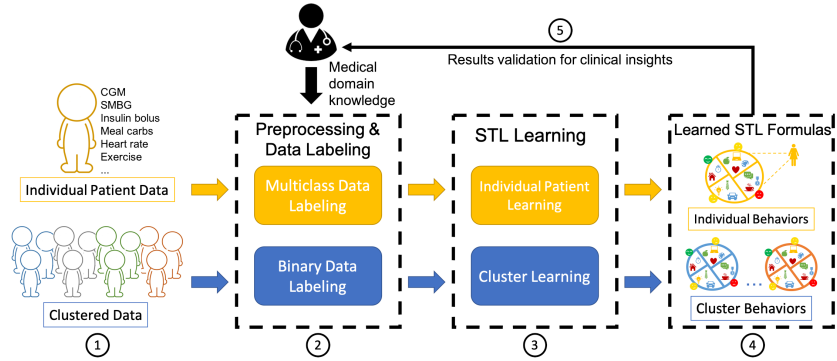}
\caption{Approach overview for learning STL formulas representing individual- (top yellow flowchart) and population-level (bottom blue flowchart) patient behavior patterns.} 
\label{fig-methods}
\vspace{-15pt}
\end{figure}

The approach of learning for individual behaviors is shown in \figref{fig-methods} in the top yellow flowchart. We first pre-processed the data, and then added a multi-class labeling mechanism for our unlabeled patient data using CGM time in range, based on medical domain knowledge. Next, we fed our data and labels into our STL learning tool, to output STL formulas that classify specific patient behaviors. Finally, our results were validated for clinical insights. Each of these steps is explained in greater detail below.


\startpara{Data Pre-processing.} As our clinical data was messy and sampled at different rates, 
the first step in our methodology was to pre-process the data. We combined all data variables into a single file, and aligned them on a five minute sampling rate, to match the set sampling rate of the CGM. The variables HR and steps were sampled at more frequent rates (data was recorded a couple times per minute,) and we used a sliding average to compute the HR value, and summed the total steps in the time frame to align with each five minute interval. In addition, we added a detector to indicate when patients were exercising. For the purposes of our approach, we determined that a patient was exercising when the Fitbit Activity Level was $\geq$ 3,
and/or when the patient had $\geq$ 3000 steps in 30 minutes, following approaches used to detect exercise in~\cite{bumgardner,marshall2009translating}. Finally, the data was layered into one hour time chunks to be fed into our STL Learning algorithm. We choose a one hour time chunk such that we would have enough data points (12 points) per layer  
for interesting learning to happen, but also small enough to provide detailed granularity within each patient's data.


\startpara{Labeling.} Next, we added labels for each one hour time chunk by hand. We used CGM Time in Range~\cite{fabris2016risk,kovatchev2017metrics}---the percentage of time a patient spends in a well controlled blood glucose range (between 70 and 180 mg/dL)---as our labeling mechanism. This metric is commonly used by clinicians to determine how well controlled patients are, and as such served as an appropriate labeling mechanism for our data. We developed 4 sets of labels based on the total percentage of time the patient was in a well-controlled range: 100\%, 75-99\%, 50-74\%, and $<$50\% time in range. These thresholds were chosen based on clinical advice, and to evenly stratify the labels across our data points.  
Since our STL Learning algorithm cannot handle mutli-class labels, we had to create 4 label sets for each of these classes with binary indicators. 
In essence, for each label set, the hour chunk of data was given a positive label if it met the correct time in range (i.e. 100\%) and a negative label if not. An example labeling scheme for a patient is shown in \figref{fig-pt-labels}. For instance, for the first label class, for each one hour layer, if 100\% of the CGM data points are in well controlled ranges then the label is a +1, and if it is anything else, then it is a -1 label. This is repeated for every hour chunk of the patient's data. For the second labeling class (75-99\% label,) a +1 label is given if 75-99\% of the CGM data points are in well controlled ranges for the hour time chunk, and -1 label if not, and so on for the rest of the data and label classes.

\startpara{STL Learning \& Validation.} Once we had developed our four labeling classes, we fed our dataset and each of the four labeling sets into the STL learning tool~\cite{nenzi2018robust} described in \sectref{sec-prelim}. For example, we fed the dataset with our 100\% time in range labels, then with our 75-99\% labels, etc. The tool works by generating formulas and picking the set that best separates our two classes (the positive and negative labelled classes). The tool then outputs these sets of formulas with the accuracy and misclassification rate (MCR). We define accuracy as 
$\frac{\mbox{\small True Positives} + \mbox{\small True Negatives}}{\mbox{\small Total}}$ 
and MCR as 1 - accuracy. In our case, we end up with 4 different final formula sets for each of our labeling classes. These formulas represent specific \emph{rules} that classify particular patient behaviors with positive and negative labels. 
A formula is considered a good candidate for characterizing data with a given label if it separates the +1 and -1 classes with a high accuracy and low MCR.
For instance, if we are classifying data using the 100\% labels, a returned formula is good if it has a high percentage of data instances correctly classified in the positive label (+1, meaning they belong to the 100\% class.)


\subsection{STL Learning for Clustered Population Behaviors}\label{sec-methods-pop}
The approach for learning population behaviors is shown in \figref{fig-methods} in the bottom blue flowchart.
First, we cluster the patient data into four population groups based on the overall percentage of time patients are in a well controlled CGM range. 
Next, the data is pre-processed and labeled, 
and the STL learning tool is used to learn formulas representative of our patient clusters. Four sets of formulas (for each of our clusters) are outputted 
and our results are validated for clinical insights at a population level.

\begin{figure}[t]
\centering
\subfigure[][]{%
\label{fig-cluster-graph}%
\includegraphics[scale = 0.29]{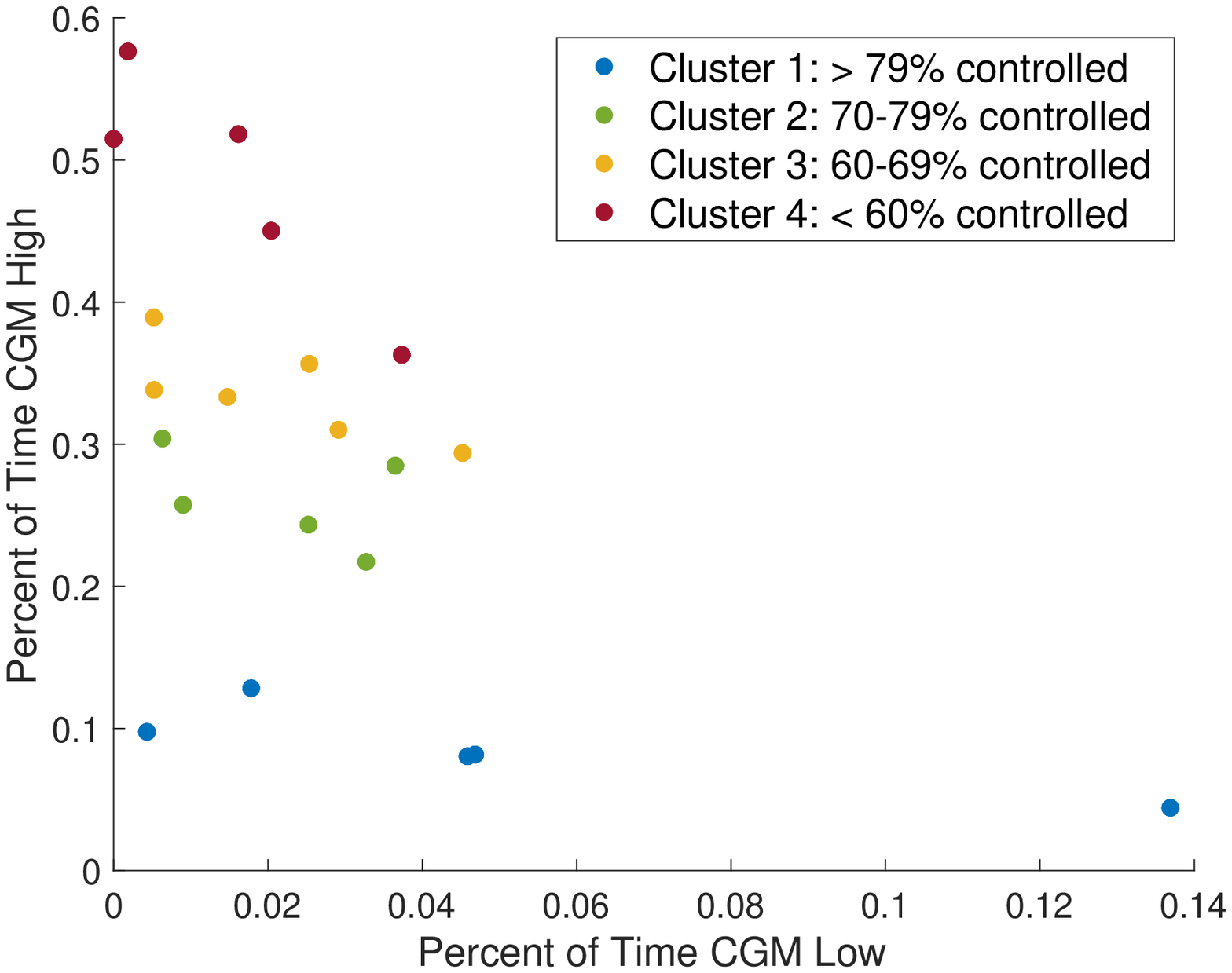}}%
\hspace{1pt}%
\subfigure[][]{%
\label{fig-pop-beh-discriminate}%
\includegraphics[scale = 0.29]{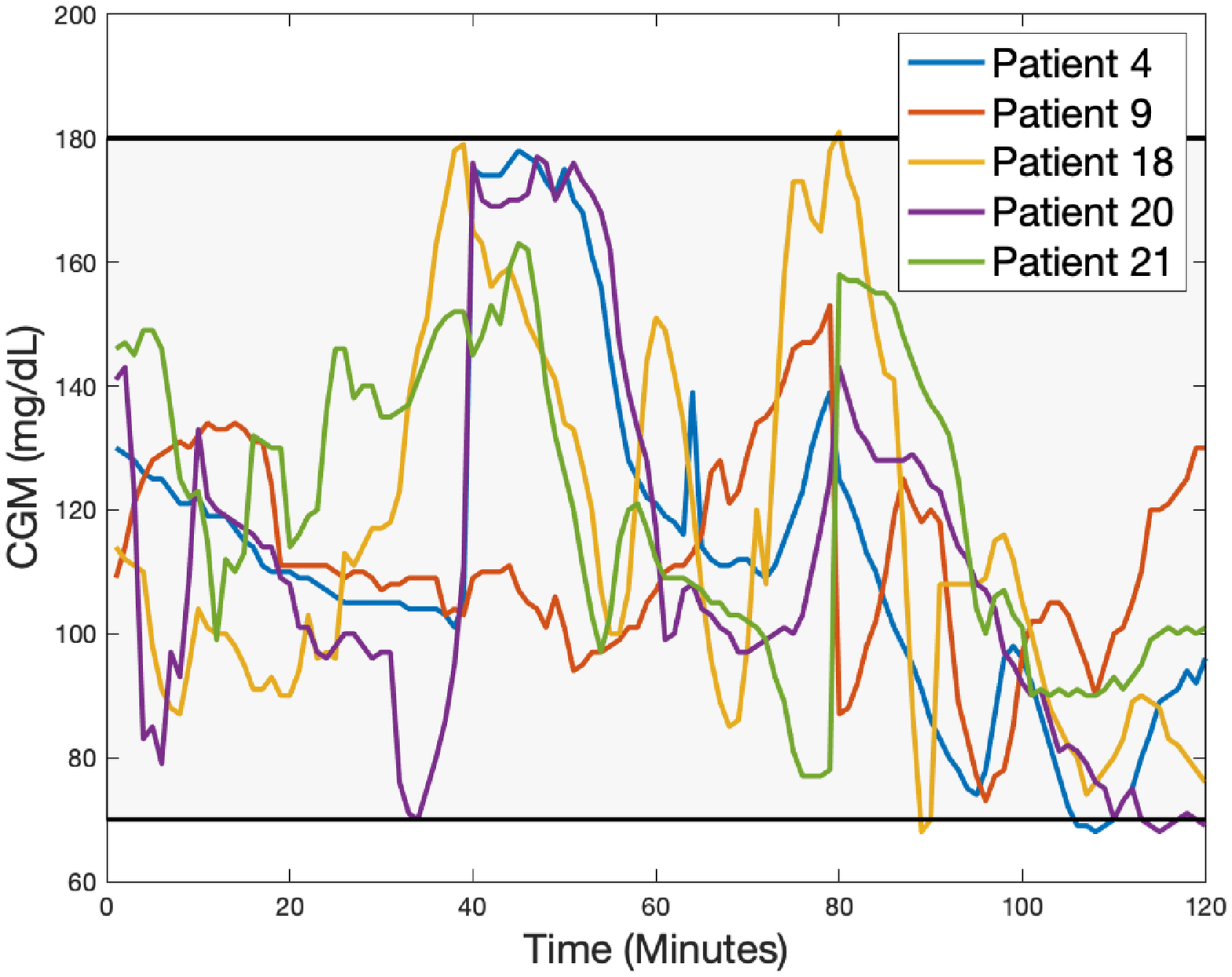}}%
\caption{\textbf{\subref{fig-cluster-graph}} Clusters of Patient Data plotted for percentage of time patients are in a high blood glucose range ($>$180 mg/dL) vs in a low blood glucose range ($<$70 mg/dL), and 
\textbf{\subref{fig-pop-beh-discriminate}} Sample patient trajectories of Cluster 1 (well controlled $>$79\% of the time).}%
\label{fig-pop-method-imgs}%
\vspace{-15pt}
\end{figure}

\startpara{Clustering.} 
The first thing we did was divide our patient data into clusters based on how well controlled they were for the entire time period of data (approx. 2 months per patient), based on the average CGM time in range. 
We had 4 clusters, grouped by best controlled patients to worst: 
Cluster 1 had patients that were well controlled $>$ 79\% of the time, Cluster 2 had patients that were well controlled 70-79\% of the time, Cluster 3 had patients that were well controlled 60-69\% of the time, and Cluster 4 had patients that were well controlled $<$60\% of the time. We clustered patients in this way to ensure a relatively even distribution of patients per cluster ($\sim$5 patients per cluster). \figref{fig-cluster-graph} shows a plot of the different patient clusters with the percentage of time their blood glucose is high ($>180$ mg/dL) vs the percentage of time their blood glucose is low ($<$70 mg/dL). We then pre-processed and chopped the data into 1 hour time chunks, following the same methodology used for individual patients.


\startpara{Labeling.} Next, we labeled each of our four clusters using a binary methodology: For each hour time chunk, if patients were 75-100\% controlled, a positive label was added (+1), and if they were $<$75\% a negative label was added (-1). It is important to note here that although patients were clustered based on their overall average time in range (i.e. Cluster 1 is for patients who had $>$79\% average time in range,) the patients within each cluster are not always in those set ranges, and there may be periods where they are more or less controlled than their average. As a result, it is necessary to label each time chunk individually based on the actual percentage of time they are in range for that \emph{specific time chunk}. For each cluster we generated one labeling set.

\startpara{STL Learning \& Validation.} We then fed each cluster and its binary label set into the STL Learning tool individually, to output four formula sets, representative of each of the clusters patients' behaviors with accuracy and MCR metrics. Similar to the STL learning for individual patients, our outputted formulas are representative of \emph{rules} that characterize the population level behaviors of the cluster. 
For example, \figref{fig-pop-beh-discriminate} shows some sample CGM trajectories of Cluster 1 patients, and the learned STL formula that characterizes these trajectories is
$\square(cgm \geq 70 \land cgm \leq 180)$.


%% file: results-individual.tex
\section{Learning Results for Individual Behaviors}\label{sec-res-individual}
In the following, we present our key findings about 
\emph{personalized} STL formulas (rules) learned from individual patients' data using the methodology described in \sectref{sec-methods-individual}.

\subsection{Personalized Bounds from Repeated Rules}

\begin{figure}[t]
\centering
\subfigure[][]{%
\label{fig-temp-bnds-a}%
\includegraphics[scale = 0.3]{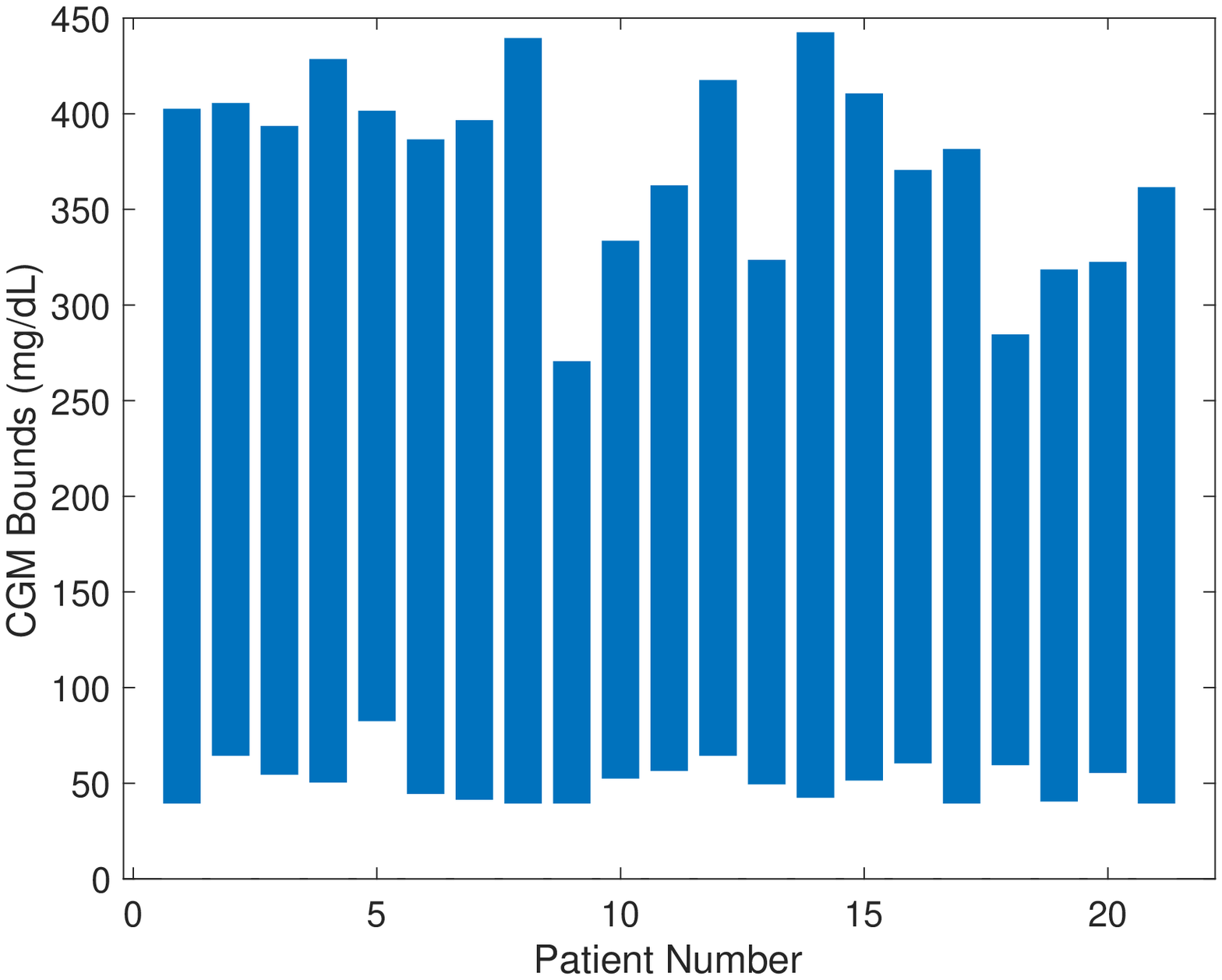}}%
\hspace{0.5pt}%
\subfigure[][]{%
\label{fig-temp-bnds-b}%
\includegraphics[scale = 0.3]{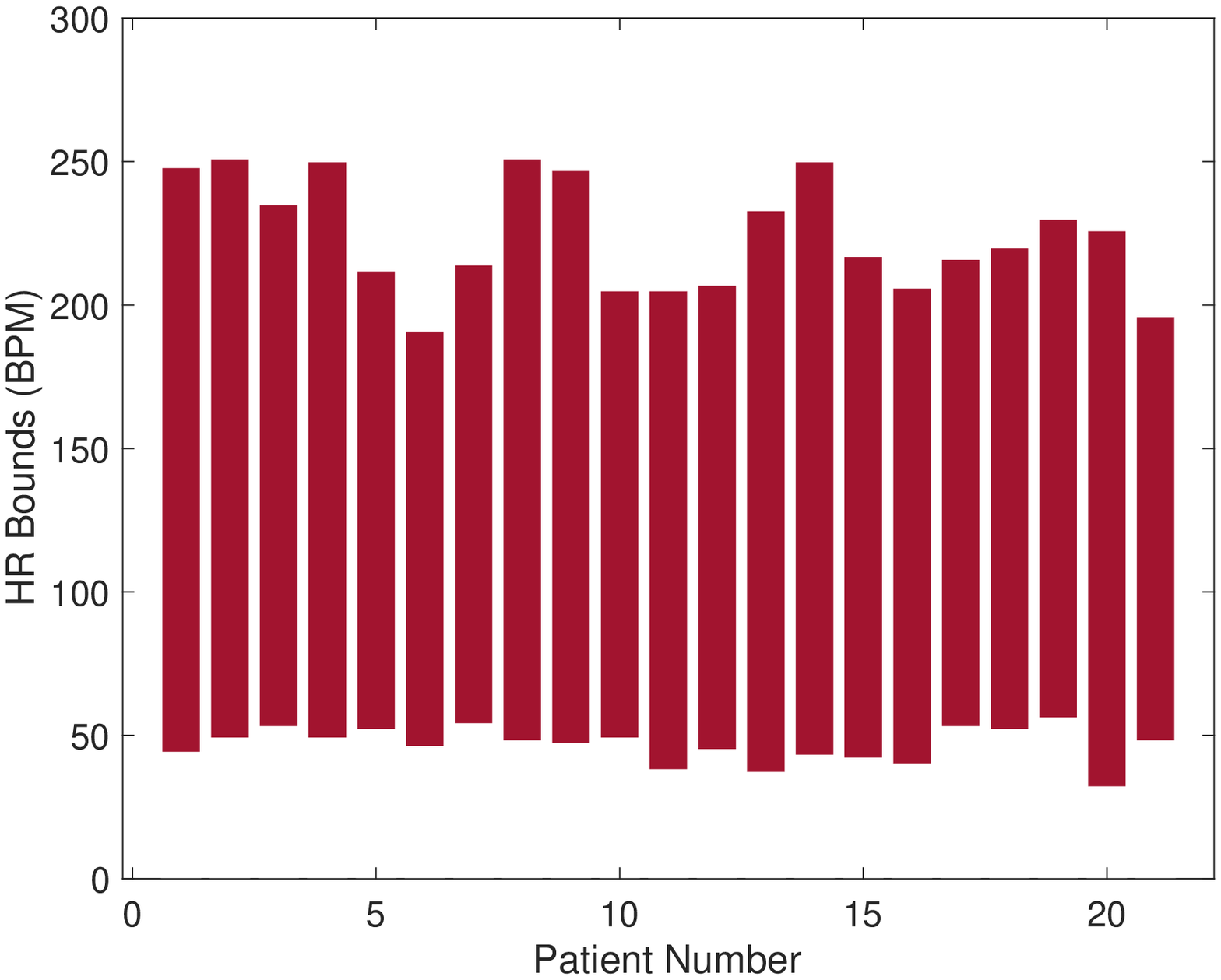}} \\
\subfigure[][]{%
\label{fig-temp-bnds-c}%
\includegraphics[scale = 0.3]{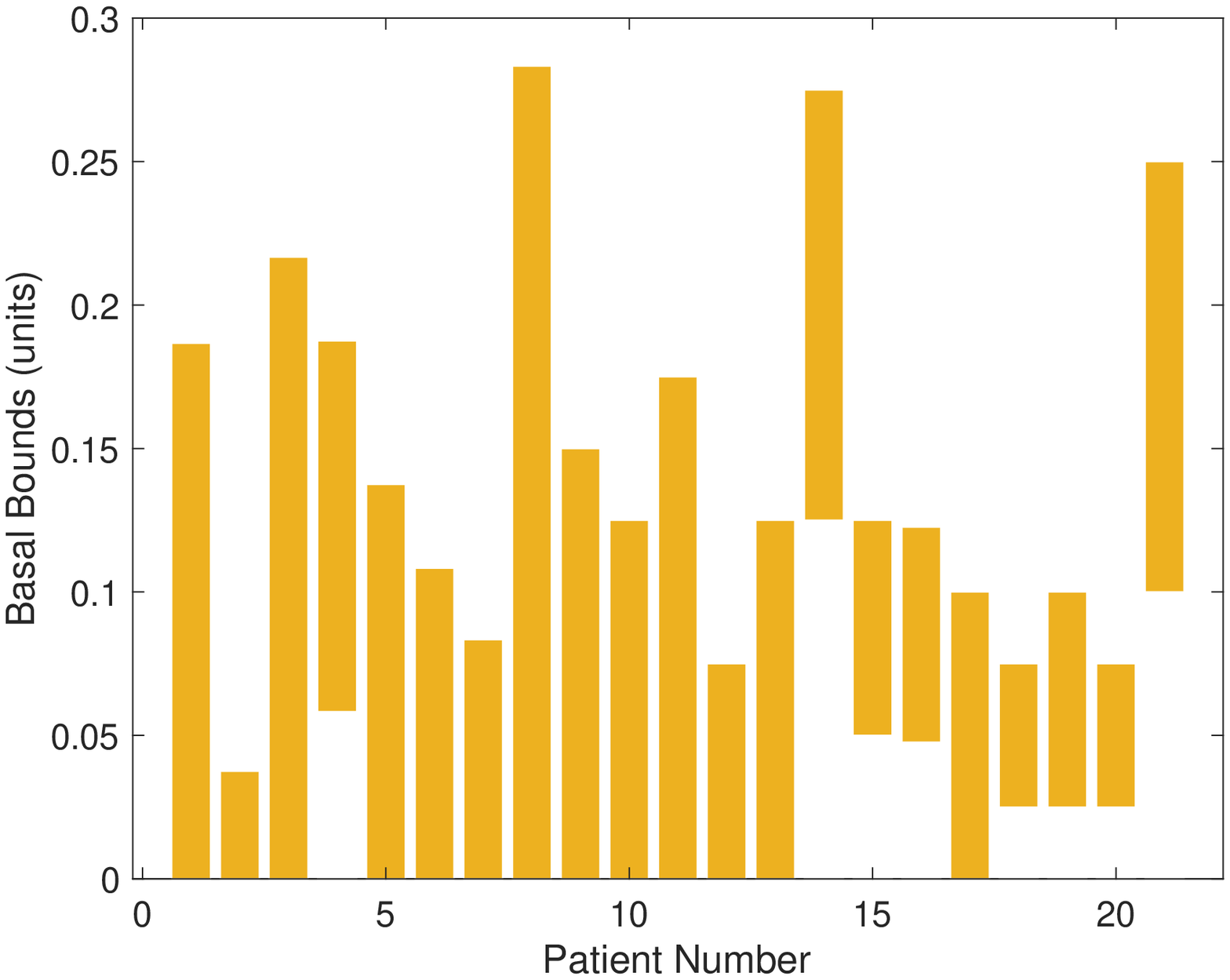}}%
\hspace{0.5pt}%
\subfigure[][]{%
\label{fig-temp-bnds-d}%
\includegraphics[scale = 0.3]{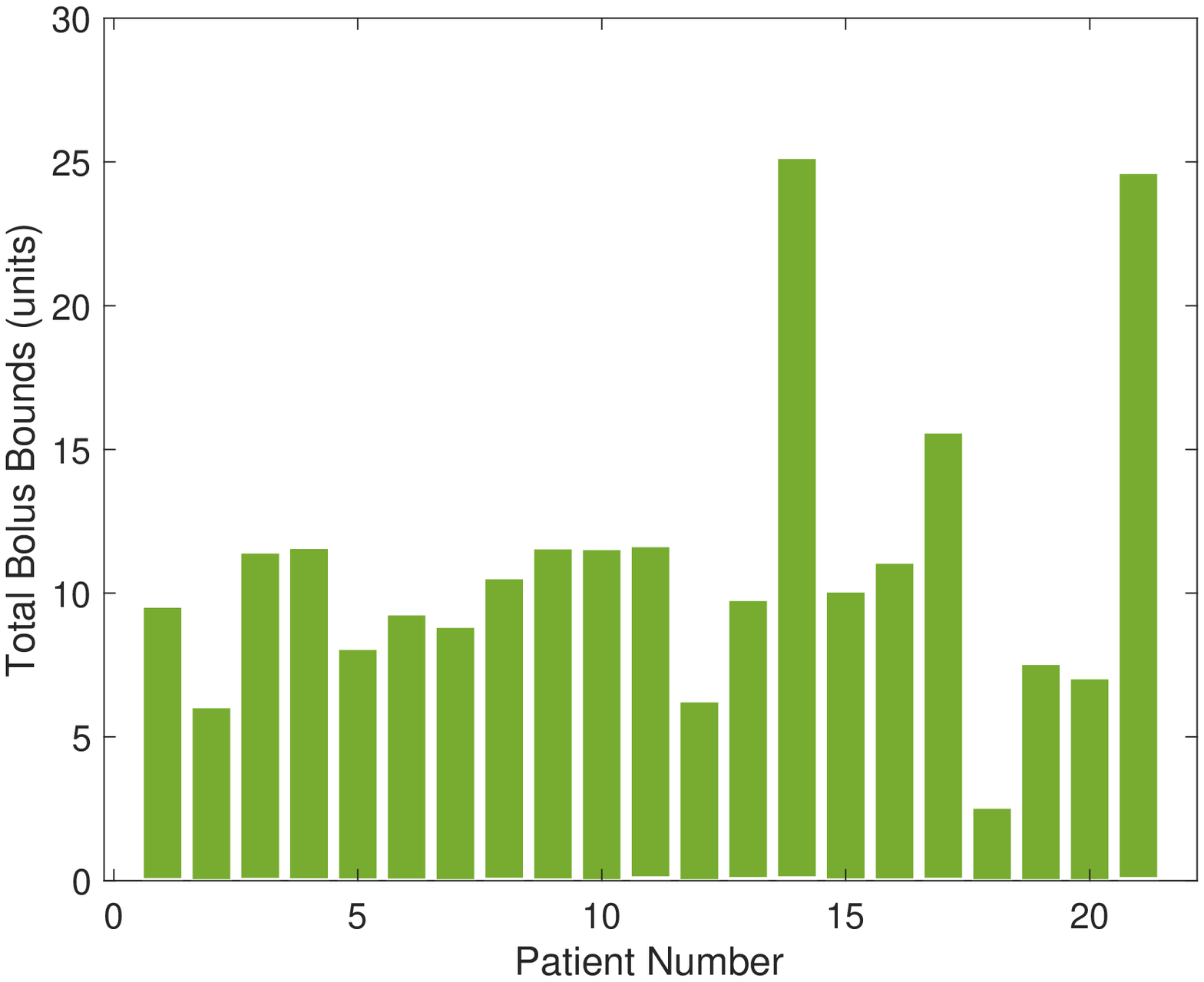}}%
\caption{Individual patient bounds for CGM\subref{fig-temp-bnds-a}, Heart Rate\subref{fig-temp-bnds-b}, Basal Bolus\subref{fig-temp-bnds-c} \& Total Bolus\subref{fig-temp-bnds-d} found from Repeated Rules (see Rule~\ref{eq-temp-rules-cgm}) for each patient's 2-months of data.}%
\label{fig-temp-bnds}%
\vspace{-15pt}
\end{figure}

One of the first interesting things we found were repeating formulas for different patients that had the same STL formula structure, but different personalized parameters, representative of patient bounds for specific variables. We identified 
repeated rules for CGM, HR, basal bolus and total bolus.
The structure of such rules is shown as follows, 
    \begin{equation} \label{eq-temp-rules-cgm}
        \varphi = \square_{[0, 1]}(x \geq \alpha \land x \leq \beta)
    \end{equation} 
where the time interval bound is within 1 hour, $x$ is the signal variable (e.g., cgm), and $\alpha$ and $\beta$ are parametric lower and upper bounds of the signal variable.
\figref{fig-temp-bnds} shows personalized parameter values learned for different patients. Since these rules encompass the range of specific bounds patients have for different data variables (i.e. CGM and HR), they are good classifiers for our data, and therefore show up repeatedly for each patient. This is supported by the fact that these rules generally (with the exception of HR bounds), have high accuracy rates (see \tabref{table-mcr-temp-rules} in the Appendix). 
Since HR is highly variable even just for an individual patient, it is not surprising that their accuracy is not extremely high. However, we include the bounds in our results for all patients, since these rules did show up repeatedly, and were accurate for some patients. We will explain the significance and use of the personalized bounds for each specific variable next.


Identifying CGM bounds as in \figref{fig-temp-bnds-a} allows for an understanding of the range of blood glucose values patients may have 
within a specific time period (in this case within a 2 month time period for our data). 
This may be relevant to note to help clinicians tailor treatment options, especially if a patient consistently has very large CGM ranges over periods of many months: 
the clinician may find it useful to find out the source of such wide variability, as well as determine other options that might help the patient reduce such large hypo- or hyper-glycemic occurrences. 
Visualizing HR bounds as shown in \figref{fig-temp-bnds-b} provides an overview of the minimum and maximum heart rate values a patient experiences. Although not clinically significant, it can be used as a quick, ballpark idea of patients' maximum heart rates, as well as their normal resting heart rates. One of the most interesting and clinically relevant bounds we are able to identify is personalized basal insulin bounds for patients, as shown in \figref{fig-temp-bnds-c}. These are very useful for determining the appropriate basal rates for individual patients. Currently, such bounds are estimated based on clinical expertise and then changed over time (using a guess-and-check method) after conferring with patients. Being able to determine the proper ranges for patients in an automated way over time is a great advantage of this approach, and can help clinicians and patients save time. 
The fourth type of bound we are able to identify is total bolus bounds, as shown in \figref{fig-temp-bnds-d}. These bolus amounts are also variable by patient, and include total meal, correction and basal bolus amounts. Although not quite as 
helpful as the 
basal bolus bounds, they still provide clinicians with an overview of the range of bolus amounts patients may have for certain time periods.

In addition, we also identified repeated rules for meal carbs and time bounds when eating occurs
, as well as exercise intensity and timing 
on a patient-by-patient basis. These rules were identified across all 4 label classes by comparing learned rules from similar time periods. For instance, we compared all of the  
rules for a single patient in the morning time (i.e. between 7:00 and 11:00) and noticed repeating rules that identified meal and exercise times for that individual patient. These rules are defined as follows,
    \begin{equation} \label{eq-temp-rule-meal-ex}
        \varphi = \Diamond_{[\alpha, \beta]}(x \leq \kappa \land y \geq \lambda)
    \end{equation} 
where $\alpha$ and $\beta$ are the time bound parameters, $x$ and $y$ are the variables the bounds are generated for (meal, HR, steps, or activity level), and $\kappa$ and $\lambda$ are 
parameters. 
As mentioned before, we do not focus on the use of rules to discriminate between different classes, but rather on the \emph{types} of behaviors we can classify within and across different label classes. In this case, these rules allow for an understanding of when  
patients are eating and exercising. 
As an example, we identified this rule 
for Patient 1 with an accuracy rate of 70.09\%, 
indicating the patient consistently eats a meal of 10 to 65 carbs between 18:01 and 19:37. 
\begin{align}
        \varphi = \Diamond_{[18:01, 19:37]}(meal \leq 65 \land meal \geq 10)
\end{align}

In another example, we identified this rule for Patient 15 (accuracy 88\%) indicating the patient consistently exercises at a moderate intensity level (Fitbit Activity Level indicator of 3 or above) between 17:59 and 19:00:
\begin{equation}
        \varphi = \Diamond_{[17:59, 19:00]}(HR \leq 212 \land activityLevel \geq 3)
\end{equation} 


\subsection{Unique Formula Relationships}
We also generated a variety of unique formulas that enable learning 
about the specific relationships between different variables (e.g. 
exercise and CGM) for individual patients. 
These relationships were identified across all our label classes, and as such may indicate 
behaviors resulting in better/worse glycemic control. However, in this section we only elucidate the \emph{types} of rules we 
identify, and we will discuss the implications for 
good and bad control in \sectref{sec-incident-good-bad-control}. 

\startpara{Meals \& CGM.} We are able to identify relationships between eating and the resulting change in blood glucose values within all four of our labeling classes. An example formula for Patient 3 showing that the CGM changes within 35min after the patient eats from the 75-99\% class is shown below (MCR = 9.09\%): 
\begin{equation}
        \varphi = ((meal \geq 1)\:\until_{[13:15, 13:50]}\:(cgm \geq 120))
\end{equation} These rules are useful to quickly understand how meals affect a patient's blood glucose and the specific time range these effects occur. 

\startpara{Meals \& Meal Bolus.} We can identify relationships for the amount of meal bolus given for various meals. For example, we find the following rule for Patient 11 (MCR 13.63\%) in the 50-74\% class, which states that they will have a meal bolus of $\leq$ to 0.8 units given a meal $\geq$ to 23 carbs:
\begin{equation}
        \varphi = \square_{[8:09, 9:09]}(mealBolus \leq 0.8 \wedge meal \geq 23)
\end{equation} These rules provide some insight into the  
bolus levels for patients based on their carb amount. This is useful to understand 
to help patients tune and identify the correct amounts of bolus they should infuse based on the carbs they eat.
\startpara{Exercise \& CGM.} Similar to meals and CGM, we can identify specific relationships about the effect exercise has on patient blood glucose levels. For example, we identify the following formula for Patient 1 (MCR = 17.143\%) in the 100\% class, which states that the patient's CGM value is greater than 120 mg/dL whenever the patient has an activity level of 3 or greater.
\begin{equation}
        \varphi = \square_{[20:31, 21:14]}(cgm \geq 120 \wedge activityLevel \geq 3)
\end{equation} These rules are useful to quickly understand how exercise may affect a patient's blood glucose and the specific time range these effects occur. 

\startpara{Eating \& Exercise.} We can also identify different instances of eating and exercise. These include eating before exercise as shown in Rule~\ref{eq-ex-meal-before} for Patient 20 (MCR = 0\%) from the $<$50\% class and eating during actual periods of exercise, as shown in Rule~\ref{eq-ex-meal-during} for Patient 21 (MCR = 5\%) from the $<$50\% class.
\begin{align}
        \varphi = ((meal \geq 1)\:\until_{[18:17,18:32]}\:(activityLevel \geq 2)) \label{eq-ex-meal-before}\\
         \varphi = \Diamond_{[12:52, 13:07]}(activityLevel \geq 2 \wedge meal \geq 10) \label{eq-ex-meal-during}
 \end{align} 
These rules are interesting to identify as they provide insights for clinicians into the strategies specific patients use to help keep their blood glucose in the proper ranges before (or during) exercise. For instance, some may eat a small snack before they begin their workout to help prevent hypoglycemia, and others may begin their workout, then realize they are becoming hypoglycemic and eat a snack during a break in the workout to prevent this.

\startpara{Exercise \& Basal Bolus.} Finally, we are able to identify basal adjustments before or during the start of exercise, such as the one shown in the rule below for Patient 11 (MCR = 7\%) from the 75-99\% class:
\begin{equation}\label{eq-ex-basal}
\varphi = ((basalBolus \leq 0.0345)\:\until_{[10:48, 11:22]}\:(activityLevel \geq 3))
\end{equation} Similar to the rules identified for eating and exercise, these rules provide insight into decisions patients make to manage their blood glucose before exercise.

\subsection{Behavioral Interventions}
We were able to generate rules that identify specific behavioral interventions patients engage in, across our four label classes. As a reminder, the focus of our approach is to characterize behaviors within labeled classes, and these rules provide insights into when patients are intervening in their T1D management, 
by double checking their blood glucose values and/or making corrections to their bolus levels. 
These rules are interesting as they indicate 
how proactive 
individuals are in monitoring and adjusting aspects of their glycemic control. Patients double check their blood glucose through a finger prick for SMBG values, and 
these formulas provide information about the circumstances  
under which patients may check their blood glucose. For instance, they may occur at regular time intervals, 
or around other events such as 
hyper- or hypo-glycemia as in Rule~\ref{eq-smbg2} (Patient 4, MCR = 0\%), or before exercise as in Rule~\ref{eq-smbg3} (Patient 8, MCR = 14.5\%). 
\begin{align}
    \varphi = \Diamond_{[14:09, 14:29]}(cgm \geq 195\:\&\: smbg \geq 200) \label{eq-smbg2}\\
    \varphi = ((smbg \geq 82)\:\until_{[10:36, 11:59]}\:(activityLevel \geq 3))\label{eq-smbg3}
\end{align}

In addition, our rules identify correction bolus times and amounts, such as those in Rule~\ref{eq-corrbolus1} (Patient 13, MCR = 12.12\%) and~\ref{eq-corrbolus2} (Patient 4, MCR = 5\%).
\begin{align}
    \varphi = ((basalBolus \leq 0.04)\:\until_{[8:15, 11:48]}\:(corrBolus \geq 0.459))\label{eq-corrbolus1}\\
    \varphi = \Diamond_{[16:58, 17:55]}(totalBolus \leq 2.105 \wedge corrBolus \geq 4.07)\label{eq-corrbolus2}
\end{align}




\subsection{Occurrences of Good and Bad Control}\label{sec-incident-good-bad-control}
Using our unique relationships, 
we were able to identify specific 
instances that patient behaviors may have resulted in good or bad control, based on which class label the rule was identified in. We identified many different types of rules classifying these 
behaviors, 
but due to space constraints we provide 6 total rules 
with their MCR in \tabref{table-incidents-good-bad}. For instance, in the case of good control we identified periods where patients were hypoglycemic and ate a meal (to raise their blood sugar,) were hyperglycemic and added a meal bolus (to lower blood sugar), and where the correct amounts of correction boluses were taken. 
For incidents of bad control, we identified periods where patients exercised but their blood glucose was too low (and no corrective actions were taken,) instances where incorrect bolus amounts for meals were taken and instances of incorrect basal or bolus adjustments. These rules are very helpful on a personalized level to help patients identify and correct behaviors that result in bad glycemic control.

\begin{table}[t]
\caption{Example Rules Capturing Good and Bad Instances of Control}\label{table-incidents-good-bad}
\centering
\begin{tabular}{|c|c|c|c|} \hline
\textbf{Class Label} & \textbf{Patient} & \textbf{Formula} & \textbf{MCR} \\ \hline
Good: 100\% & 1 & $\varphi = \Diamond_{[13:24, 15:22]}(smbgHypo \geq 1 \wedge meal \geq 49)$ & 0\% \\ \hline
Good: 75-99\% & 2 & $\varphi = \Diamond_{[12:00, 12:55]}(smbgHypo \geq 1 \wedge totalBolus \leq 7.18)$ & 0.2\% \\ \hline
Good: 100\%& 21 & $\varphi = ((activityLevel \leq 4)\:\until_{[10:36, 11:59]}\:(corrBolus \geq 5.9))$ & 5\% \\ \hline
Bad: 50-74\%& 5 & $\varphi = \Diamond_{[15:00, 17:41]}(cgm \leq 68 \wedge basalBolus \leq 0.011)$ & 13.18\% \\ \hline
Bad: $<$50\% & 7 & $\varphi = \Diamond_{[11:55, 13:02]}(activityLevel \geq 4 \wedge cgm \leq 65)$ & 1.8\% \\ \hline
Bad: $<$50\%& 15 & $\varphi = ((meal \leq 44)\:\until_{[21:09, 23:37]}\:(cgm \geq 210))$ & 6.36\% \\ \hline
\end{tabular}
\vspace{-15pt}
\end{table}

\subsection{Example Use Case}
We next present a sample use case of our learned rules. Using the formulas generated for occurrences of good and bad control, we can identify the specific basal bolus amounts appropriate for different exercise intensity levels for a specific patient (i.e. Patient 21). \tabref{table-basal-ranges} shows the minimum and maximum basal bounds for each activity level, and the Misclassification Rate for the good and bad formulas (MCR Good and MCR Bad), and the good and bad classification formulas used to derive each of the basal range bounds are shown below (the name of each $\varphi$ indicates the label class and the activity level.) 
We define the 75-99\% and 100\% labels as the ``good class" and the 50-74\% and $<$50\% labels as the ``bad class". For instance, in the first row of \tabref{table-basal-ranges} we can see that the basal range is between 0.066 and 0.072 for an activity level of 4. We reference Rule~\ref{eq-act-4-good}, that states that the basalBolus is below 0.072 units at the start of intense exercise (activity level 4,) and Rule~\ref{eq-act-4-bad}, that states that bad control occurs when exercise activity level is 4 and the basal bolus is less than 0.065 (meaning we need a higher basal rate than this for good control.) From these we can derive the basal rate bounds: an upper rate bound of 0.072 from our good classification formula, and a lower bound of 0.066 from our bad classification formula.

\begin{table}[h]
\vspace{-15pt}
\caption{Proper Basal Ranges for Exercise Intensity for Patient 21}\label{table-basal-ranges}
\centering
\begin{tabular}{|c|c|c|c|c|} \hline
\textbf{Act. Level} & \textbf{Basal Range} & \textbf{Formulas Used} & \textbf{MCR\textsubscript{Good Class}} & \textbf{MCR\textsubscript{Bad Class}}\\ \hline
4 & 0.066 - 0.072 &  \ref{eq-act-4-good}, \ref{eq-act-4-bad} & 14.84\% & 0\% \\ \hline
3 & 0.073 - 0.077 & \ref{eq-act-3-good}, \ref{eq-act-3-bad} & 16.23\% & 10.12\%\\ \hline
2 & 0.078 - 0.089 & \ref{eq-act-2-good} & 0\% & N/A\\ \hline
1 & 0.09 - 0.1 & \ref{eq-act-1-good}, \ref{eq-act-1-bad} & 26.35\% & 1.8\% \\ \hline
\end{tabular}
\vspace{-15pt}
\end{table}

\startpara{Formulas Used to Derive \tabref{table-basal-ranges}:}
\begin{align}
\vspace{-15pt}
    &\varphi_{good4} = \square_{[9:00, 11:01]}(basalBolus \leq \textbf{0.072})\until_{[9:10, 11:01]}(activityLevel \geq 4) \label{eq-act-4-good}\\
    &\varphi_{bad4} = \square_{[9:00, 11:05]}(activityLevel \geq 4 \wedge basalBolus \leq \textbf{0.065}) \label{eq-act-4-bad}\\
    &\varphi_{good3} = \square_{[9:00, 11:00]}(activityLevel \leq 3 \wedge basalBolus \leq \textbf{0.072}) \label{eq-act-3-good}\\
    &\varphi_{bad3} = \square_{[9:02, 10:59]}(activityLevel \geq 3 \wedge basalBolus \geq \textbf{0.078}) \label{eq-act-3-bad}\\
    &\varphi_{good2} = \square_{[8:58, 11:00]}(activityLevel \leq 2 \wedge basalBolus \leq \textbf{0.089})\label{eq-act-2-good}\\
    &\varphi_{good1} = \square_{[8:55, 10:57]}(activityLevel \leq 1 \wedge basalBolus \geq \textbf{0.091})\label{eq-act-1-good}\\
    &\varphi_{bad1} = \square_{[8:55, 11:05]}(basalBolus \leq \textbf{0.122})\:\until_{[9:10, 11:01]}\:(activityLevel \geq 1) \label{eq-act-1-bad}
\end{align}



%% file: results-population.tex
\section{Learning Results for Population Behaviors}\label{sec-res-pop}
We now present results of population-level patient behaviors learned using the methodology in \sectref{sec-methods-pop}. 
There are several interesting key findings. First, the most controlled patients had the most number of SMBG occurrences (double checks of their blood glucose) as shown in \tabref{table-smbg-count}. These occurrences were drawn from our STL formulas generated related to SMBG, an example of which is displayed in Rule~\ref{eq-ex-smbg-pop}. As mentioned before, Cluster 1 contains the best controlled patients, and Cluster 4 contains the worst controlled patients. This finding indicates that the best controlled patients 
double check their blood glucose much more frequently, which may result in better overall control of their T1D. This makes sense, because patients who are more actively engaged in  
verifying the status of their blood glucose (and other factors of their glycemic control,) are more proactive in making the necessary changes (i.e. 
adding a correction bolus) 
in order to ensure their blood glucose stays within the proper ranges. 
Alternatively, patients who have worse control tend to check their blood glucose values less often, meaning they may not be as aware of specific blood glucose changes 
that require some adjustment to the management of their T1D.
The following is an example SMBG rule used to derive Table~\ref{table-smbg-count} for a 24 hour time period for Cluster 1 (accuracy = 100\%):
\begin{equation}\label{eq-ex-smbg-pop}
    \varphi = \Diamond_{[12:00, 12:00]}(smbg \geq 55 \wedge cgm \leq 400)
\vspace{-1pt}
\end{equation}

\begin{table}[h]
\vspace{-18pt}
\caption{Average SMBG Count By Cluster}\label{table-smbg-count}
\centering
\begin{tabular}{|c|c|} \hline
\textbf{Cluster Number} & \textbf{Average Count of SMBG Checks}\\ \hline
1 (best controlled) & 85.00\\ \hline
2 & 68.80 \\ \hline
3 & 59.67\\ \hline
4 (worst controlled) & 50.60 \\ \hline
\end{tabular}
\vspace{-15pt}
\end{table}

Second, from our rules we identified that as we go from the best controlled cluster (Cluster 1) to the worst (Cluster 4), we have an increased count of 
correction boluses per patient. This is shown in the second column of \tabref{table-corrbolus-numAmt}, and some sample rules that we derived these values from is shown in Rule~\ref{eq-corrbolus-pop}. Moreover, not only do patients with worse control have an increased count of the correction boluses, they also have an increased average \emph{amount} of actual correction bolus units taken per correction bolus occurrence. This is shown in the third column of \tabref{table-corrbolus-numAmt}. These findings indicate that patients who have worse control tend to need to correct their bolus levels more often, and change (i.e. increase) their actual correction bolus amounts more drastically than better controlled patients. These findings also make sense, because less controlled patients may take more of a reactive approach, (e.g. they only intervene in their control when a specific incident such as 
hyper- or hypo-glycemia occurs), resulting in an increased need to correct their bolus levels, and by larger unit amounts  
at each intervention.
The following is an example Correction Bolus rule used to derive Table~\ref{table-corrbolus-numAmt} for a 24 hour time period for Cluster 4 (accuracy = 100\%):
\begin{equation}\label{eq-corrbolus-pop}
    \varphi = \Diamond_{[23:59, 23:59]}(corrBolus \geq 10)
\end{equation} 

\begin{table}[t]
\caption{Average Number and Amount of Correction Boluses By Cluster}\label{table-corrbolus-numAmt}
\centering
\begin{tabular}{|c|c|c|} \hline
\textbf{Cluster Number} & \textbf{Number of Correction Boluses} & \textbf{Correction Bolus Amount}\\ \hline
1 (best controlled) & 8.80 & 17.14\\ \hline
2 & 11.80 & 18.16\\ \hline
3 & 12.17 & 23.98\\ \hline
4 (worst controlled) & 14.80 & 32.30\\ \hline
\end{tabular}
\vspace{-15pt}
\end{table}

We were not able to identify any other specific formulas that made sense and that provided a good characterization between the clusters. Although different rules relating CGM or exercise to other components (i.e. basal bolus) were generated, these rules cannot be used for the entire cluster population. These types of formulas and their parameters should be very specific to individuals, and therefore cannot be generalized, even across a small cluster of patients. 


%% file: relatedwork.tex
\section{Related Work}\label{sec-relatedwork}

\startpara{Learning Diabetes Patient Behaviors.} A couple of works have looked at learning patient behaviors for T1D patients at a population level. Chen et al.~\cite{Chen2015} developed an ``eat, trust, check'' framework to 
model and evaluate patient insulin pump behaviors using a machine learning approach. Hoyos et al.~\cite{hoyos2018population} used an incremental learning approach to infer the behavior of autonomous glucose measurements and parameters for population groups of T1D patients. In addition, Cameron et al.~\cite{cameron2012extended} developed a model predictive controller for regulating blood glucose based on cgm readings and meal behaviors, and Paoletti et al.~\cite{paoletti2017data} presented a model predictive controller to administer insulin based on patient behavior (i.e. meal and exercise events). These approaches 
do not include behavior types beyond meals/exercise as our approach does (such as our SMBG checks) and do not employ STL Learning, so they 
are not able to express the range of different behaviors and 
personalized level of formulas 
that our 
methodology can. Moreover, Chatterjee et al.~\cite{chatterjee2018designing} designed a sensor-based at home system for T1D patients that records patient activity throughout the day  
to promote patient behavior change. 
This approach 
provides alerts about more high-level activities (eating and sedentary behavior), and does not provide as specific of information (such as about behavioral interventions related to SMBG) as in our approach. 

\startpara{STL Learning for Behavior Detection.} In terms of STL Learning, a variety of papers have developed new methodologies to learn STL formula structures and their parameters for anomaly detection and behavior identification in applications such as naval surveillance and medical contexts. Kong et al.~\cite{Kong2017} developed an offline supervised learning approach that uses machine learning to detect anomalous and normal behaviors. Formula structures and parameters are synthesized using a gradient descent optimization guided by robustness and hinge loss functions in their machine learning algorithm. 
This work suffers from a long computational complexity, due to the time needed to optimize for the graph structure and a lack of explainability due to the ML algorithm.
Our approach is explainable and facilitates greater clinician trust in the outcome of our results. 
In addition, Klimek~\cite{Klimek2016} and Bombara et al.~\cite{Bombara} used tree structures to generate their STL formulas and parameters. Klimek employed an online learning approach in which graph models were used to  
reason about objects and events, and logical truth trees were outputted to represent the 
formulas 
and their behavioral meanings. Bombara et al. use a decision tree framework and a misclassification rate optimization method to build binary decision trees representative of STL formulas and their parameters to categorize anomalous vs normal behaviors. The strict structure of the tree algorithms imposes some restrictions on the flexibility and diverse types of STL formula structures that can be outputted. As a result, these structures are not 
optimal choices for T1D patient data, as they 
lack the expressivity needed to classify diverse 
patient behaviors. Moreover, the formulas generated from the decision tree are long and 
not very human readable.

%% file: conclusion.tex
\section{Discussion \& Conclusion}\label{sec-conclusion}

\startpara{Conclusion.} In this paper, we presented an approach to learn STL formulas that characterize individual- and population-level T1D patient behaviors with varying glycemic control and applied it to a clinical dataset with 21 T1D patients' data. 
Our learning results provide some clinically-relevant insights for clinicians and patients to develop behavioral change strategies to improve glycemic control.

\startpara{Tool Limitations.} Our results are constrained by the limitations of the STL learning tool~\cite{nenzi2018robust} in several ways. First, our patient data contain many null values. For example, patients only eat at discrete time periods, and times the patient was not eating were null. However, the tool cannot handle null values, so we had to fill all of these instances with zeros: this changes the semantic meaning of the data points, and may cause a bias in how the formula parameters are being generated (for instance, when data points are averaged to get specific parameter bounds). Second, since the tool cannot handle multi-class classification, we had to use four different sets of labels with binary indicators to cover our different classes. This may have caused some overlap in our resulting formulas. In addition, since the tool relies on a supervised classification approach, we had to supply labels to guide the learning. However, this may have resulted in missing some behavior sets that still have an effect on T1D glycemic control (but may not have a direct relationship with CGM time in range). Moreover, the tool relied on having an evenly split distribution of data labels, which proved challenging for our unevenly distributed patient data. Finally, the tool can only learn from raw data streams for short time periods, (and can not, for instance, calculate CGM rate of change or other advanced relationships), and as such we were only able to learn fairly simple rules for short time chunks. As a result, we were unable to study longer term T1D effects (e.g. multiple hour meal-bolus relationships). 
\startpara{Future Work.} We would like to address the limitations and improve upon the capabilities of the tool, as well as integrate our patient behavior identification approach into a closed loop feedback system (e.g., implemented in a smartphone application or other wearable), which will provide real-time feedback about behaviors that have negative impact on glycemic control.

